\documentclass[aps,prb,reprint,superscriptaddress,nofootinbib,floatfix]{revtex4-2}
\usepackage{amsmath,amssymb,bm,graphicx,microtype}
\usepackage[hidelinks]{hyperref}

\begin{document}
\title{Odd-parity magnon spin polarization from chiral four-spin exchange in a honeycomb antiferromagnet}
\author{B. Murzaliev}
\affiliation{Materials and Structural Analysis Division, ThermoFisher Scientific, Achtseweg Noord 5, 5651 GG, Eindhoven, The Netherlands}
\author{M. Titov}
\affiliation{Institute for Molecules and Materials, Radboud University, 6525 AJ Nijmegen, The Netherlands}

\begin{abstract}
Chiral four-spin exchange generates a momentum-dependent magnon spin in a compensated honeycomb antiferromagnet at zero magnetic field. For a N\'eel vector tilted by an angle $\theta_0$ from the layer normal, the interaction contributes to intersublattice magnon pairing already at harmonic order, with an amplitude proportional to $\cos\theta_0\sin^2\theta_0$. Biaxial anisotropy mixes the two longitudinal-spin sectors, producing nonquantized mode spins that are odd in momentum, while a sublattice-exchange relation keeps the magnon energies reciprocal. A temperature gradient then induces a spin density along the N\'eel axis through a redistribution of magnon occupations. Reversing the four-spin coupling reverses this spin density and the mode spins without changing the magnon energies. MnPS$_3$ motivates the magnetic geometry; realizing the interaction requires a structural modification compatible with the symmetry assumptions of the model.
\end{abstract}
\maketitle

\section{Introduction}

The ordered moments of a compensated antiferromagnet cancel, but its magnons need not be spinless: An individual mode can carry a spin whose magnitude and direction depend on momentum \cite{Okuma2017}. When spin-rotation symmetry about the ordering axis is broken, this spin is not an integer multiple of $\hbar$ \cite{Kamra2017}. If modes at opposite momenta have equal energies and opposite spin, the magnon spin is an odd function of momentum and cancels in equilibrium. A temperature gradient populates the counterpropagating modes unequally and leaves a net spin density --- the magnon thermal Edelstein effect \cite{Li2020,Zhang2020}, distinct from the transverse spin current of the magnon spin Nernst effect \cite{Cheng2016,Zyuzin2016}. The question is which microscopic interactions produce such odd-parity magnon spin while keeping the spectrum reciprocal.

One answer is Dzyaloshinskii--Moriya exchange \cite{Dzyaloshinsky1958,Moriya1960}. In a collinear antiferromagnet it shifts the magnon bands of opposite spin in opposite directions in momentum, and a transverse anisotropy that mixes the two bands lets their spin vary smoothly with momentum \cite{Zhang2020}. Related odd-parity magnon effects arise from intrasublattice Dzyaloshinskii--Moriya exchange in collinear multiferroics \cite{Du2026} and from noncoplanar magnetic order \cite{Neumann2026}. Chiral exchange is not restricted to bilinear couplings. Higher-order expansions of electronic models generate biquadratic and multispin interactions \cite{Hoffmann2020}, which affect the magnetic properties of two-dimensional magnets \cite{Kartsev2020}, and spin-orbit coupling makes some of them chiral \cite{Brinker2019,Brinker2020}. In two-dimensional ferromagnets with trigonal prismatic symmetry, one such interaction --- quartic asymmetric exchange --- modifies both magnetic textures and magnon spectra \cite{Rakhmanova2022,Ado2022}.

Here we show that quartic asymmetric exchange, a chiral four-spin interaction on the honeycomb lattice, produces odd-parity magnon spin in a compensated antiferromagnet already at harmonic order. In a honeycomb ferrimagnet the same interaction acts through nonlinear magnon processes, which offer a route to its optical detection \cite{Murzaliev2025}. In the antiferromagnet, a N\'eel vector tilted by an angle $\theta_0$ from the layer normal converts it into a pairing of magnons on opposite sublattices whose leading long-wavelength term is odd in momentum, with a strength $\propto\cos\theta_0\sin^2\theta_0$ --- the same angular factor that sets the spiral wave vector in the ferromagnetic case \cite{Ado2022}. Biaxial anisotropy (unequal anisotropies along the two transverse axes) mixes the magnon modes of opposite spin, and a sublattice-exchange symmetry keeps their energies reciprocal while making their spin odd in momentum. A temperature gradient then produces a spin accumulation along the N\'eel axis that changes sign with the four-spin coupling and, for a constant relaxation time, vanishes without the anisotropy.

The magnetic geometry is motivated by MnPS$_3$, a layered honeycomb antiferromagnet of Mn$^{2+}$ moments \cite{Joy1992}, whose spin waves and critical behavior have been characterized by neutron scattering \cite{Wildes1998,Wildes2006} and whose order persists down to the bilayer \cite{Kim2019}. Its ordered moments are tilted by about $8^\circ$ from the layer normal \cite{Ressouche2010}, so the tilt required for the pairing is present. Neutron spectroscopy finds no significant difference between the mean magnon energies at opposite zone corners \cite{Wildes2021}. Comparing energies at opposite momenta does not reveal this mechanism, because the dispersion remains reciprocal. At fixed coupling magnitude, reversing the coupling changes the mode spin but leaves the energies unchanged. The four-spin interaction is odd under inversion and therefore forbidden in pristine, centrosymmetric MnPS$_3$; we consider a structurally modified honeycomb antiferromagnet whose symmetry permits it (see Sec.~\ref{sec:conclusion}).

\section{Model}

\begin{figure}[t]
\centering
\includegraphics[width=\columnwidth]{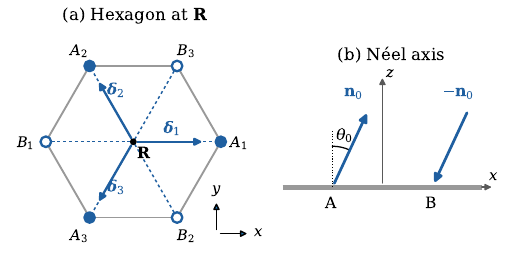}
\caption{
Geometry of the model.
(a) Hexagon centered at $\bm R$. The sites $A_\alpha$ (filled) at $\bm R+\bm\delta_\alpha$ and
$B_\alpha$ (open) at $\bm R-\bm\delta_\alpha$ lie on a common line through $\bm R$ (dashed). The
four-spin interaction~\eqref{eq:hchi} projects both spins of the pair onto the same vector
$\bm\delta_\alpha$.
(b) Side view: the N\'eel axis $\bm n_0$ lies in the $xz$ plane, tilted by $\theta_0$ from
the layer normal $z$ toward the bond $\bm\delta_1$; the $B$ moments point along $-\bm n_0$.
The tilt is exaggerated; $\theta_0=8^\circ$ in our calculations.
}
\label{fig:geometry}
\end{figure}

We consider spins of equal length $S$ on the two sublattices of a honeycomb lattice at zero magnetic field. Antiferromagnetic nearest-neighbor exchange favors antiparallel moments, and magnetic anisotropy selects their common axis, which we take in the $xz$ plane (Fig.~\ref{fig:geometry}), $\bm n_0=(\sin\theta_0,0,\cos\theta_0)$, with $z$ normal to the layer. The transverse unit vectors are $\bm e_\theta=(\cos\theta_0,0,-\sin\theta_0)$ and $\bm e_\varphi=(0,1,0)$. The reference Hamiltonian is
\begin{align}
\mathcal H_0={}&J\sum_{\langle ij\rangle}
\bm S_{iA}\cdot\bm S_{jB}
\nonumber\\
&+\sum_{i,\eta=A,B}\left[
K_\theta\left(\frac{\bm S_{i\eta}\cdot\bm e_\theta}{S}\right)^2
+K_\varphi\left(\frac{\bm S_{i\eta}\cdot\bm e_\varphi}{S}\right)^2
\right],
\label{eq:model}
\end{align}
with dimensionless spin operators and energies $J,K_\theta,K_\varphi>0$. Here $i$ labels the unit cells, $\eta=A,B$ the sublattices, and $\langle ij\rangle$ the nearest-neighbor bonds. Its classical minimum is the compensated N\'eel state $\bm S_{iA}=S\bm n_0$, $\bm S_{iB}=-S\bm n_0$. Unequal transverse anisotropies, $K_\theta\neq K_\varphi$, break spin-rotation symmetry about $\bm n_0$ and allow magnon modes of opposite spin to mix \cite{Kamra2020}.

To this we add a chiral four-spin interaction acting on individual hexagons \cite{Murzaliev2025}. For a hexagon centered at $\bm R$, the $A$ and $B$ vertices lie at $\bm R+\bm\delta_\alpha$ and $\bm R-\bm\delta_\alpha$, respectively, with $\bm\delta_1=a(1,0,0)$, $\bm\delta_{2,3}=a(-1/2,\pm\sqrt{3}/2,0)$, and $a$ the nearest-neighbor distance. Denote by $p_{\eta\alpha}=\bm\delta_\alpha\cdot\bm S_{\eta\alpha}/S$ the projection of the spin $\bm S_{\eta\alpha}$ at vertex $\bm R\pm\bm\delta_\alpha$, with the same projection vector on both sublattices and the hexagon label left implicit [Fig.~\ref{fig:geometry}(a)]. The vectors $\bm\delta_\alpha$ have units of length, whereas the spin operators and $S$ are dimensionless. Thus $p_{\eta\alpha}$ has units of length and the factor $a^{-4}$ makes $A_D$ an energy. The interaction is
\begin{equation}
\mathcal H_\chi=\frac{A_D}{a^4}\sum\nolimits_{\bm R}[\textstyle\prod_{\alpha}p_{A\alpha}\sum_{\beta}p_{B\beta}
-\prod_{\alpha}p_{B\alpha}\sum_{\beta}p_{A\beta}],
\label{eq:hchi}
\end{equation}
with $\alpha,\beta=1,2,3$. Each term couples three spins on one sublattice to a fourth spin on the other. Structural inversion exchanges the sublattices and reverses the sign of $\mathcal H_\chi$, whereas time reversal leaves it unchanged; we assume a crystal environment whose symmetry permits the interaction.

The four-spin interaction leaves the uniform N\'eel configuration stationary. Because $\sum_\alpha\bm\delta_\alpha=0$, uniform moments give $\sum_\alpha p_{\eta\alpha}=0$, so the classical energy of $\mathcal H_\chi$ vanishes. Its first-order variation also cancels after summing over hexagons in the bulk \cite{SupplementalMaterial}, so the spin-wave expansion contains no linear terms and the leading effect of $\mathcal H_\chi$ is quadratic in the fluctuations.

We introduce Holstein--Primakoff bosons $a$ and $b$ for the $A$ and $B$ sublattices \cite{Holstein1940}, with Fourier components $a_{\bm q}$ and $b_{\bm q}$ defined by the phases $e^{i\bm q\cdot\bm r}$ of the physical site positions $\bm r$. At quadratic order the number-conserving contributions from $\mathcal H_\chi$ cancel, and only intersublattice pair creation and annihilation remain,
\begin{equation}
\mathcal H_\chi^{(2)}
=-\frac{1}{S}\sum_{\bm q}
\left[
\Gamma_{\chi,\bm q}\,a_{\bm q}b_{-\bm q}
+\mathrm{H.c.}
\right],
\label{eq:pair}
\end{equation}
with, for the N\'eel axis chosen above and without a small-momentum expansion,
\begin{equation}
\Gamma_{\chi,\bm q}
=-\frac{3\sqrt{3}}{4}A_D
\cos\theta_0\sin^2\theta_0\,
e^{iaq_x/2}
\sin\frac{\sqrt{3}aq_y}{2}.
\label{eq:gammaexact}
\end{equation}
At this order two of the four spin projections take their values in the ordered background, while the other two describe transverse fluctuations on opposite sublattices. The angular factor $\cos\theta_0\sin^2\theta_0$ shows that the pairing vanishes for moments normal to the layer ($\theta_0=0$) or lying in it ($\theta_0=\pi/2$). The exact lattice relation is $\Gamma_{\chi,-\bm q}=-\Gamma_{\chi,\bm q}^*$: its real part is odd and its imaginary part even in momentum, and the leading long-wavelength term is real and linear in $q_y$. The full expansion, Eqs.~(S10)--(S12), is given in the Supplemental Material \cite{SupplementalMaterial}.

To be specific, we take the spin $S=5/2$ of Mn$^{2+}$ and the tilt $\theta_0=8^\circ$ of MnPS$_3$ \cite{Ressouche2010}, together with $A_D=0.05J$ and transverse anisotropies $K_\theta=0.02J$, $K_\varphi=0.01J$. The quadratic Hamiltonian is then positive definite throughout the Brillouin zone \cite{SupplementalMaterial}, so the compensated state is locally stable within the harmonic approximation.

\begin{figure*}[t]
\centering
\includegraphics[width=\textwidth]{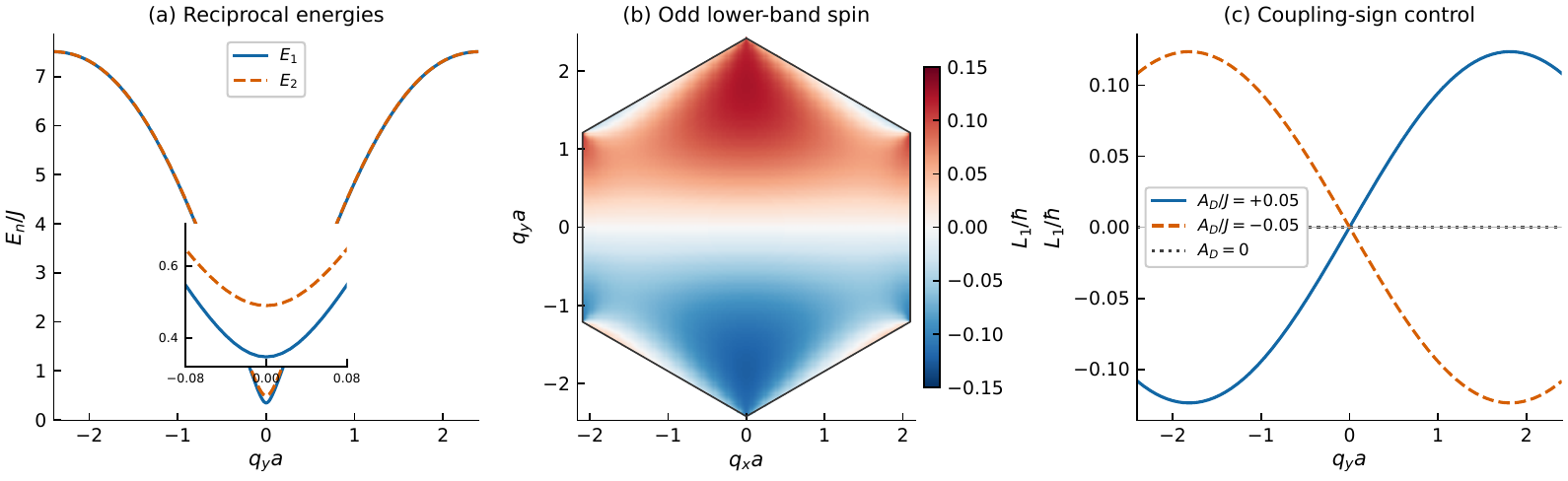}
\caption{
Reciprocal magnon energies and odd-parity magnon spin.
(a) Energies $E_{1\bm q}$ (solid) and $E_{2\bm q}$ (dashed)
of the two magnon bands along $q_x=0$. The bands are indistinguishable on this
scale; the inset resolves the gaps $0.35J$ and $0.49J$ at the zone center.
(b) Spin $L_{1\bm q}$ along the N\'eel axis, Eq.~\eqref{eq:spin},
carried by the lower band in the first Brillouin zone. The spin is odd under
$\bm q\to-\bm q$ and vanishes on the line $q_y=0$, where the chiral pairing
\eqref{eq:gammaexact} vanishes.
(c) $L_{1\bm q}$ along $q_x=0$ for $A_D/J=+0.05$ (solid),
$-0.05$ (dashed), and $0$ (dotted): reversing the four-spin coupling reverses the spin.
Parameters: $S=5/2$, $\theta_0=8^\circ$,
$K_\theta=0.02J$, $K_\varphi=0.01J$, and $A_D=0.05J$ in panels (a,b).
At the degenerate zone corners $K$ and $K'$ the spin of an individual band
depends on the choice of basis \cite{SupplementalMaterial}.
}
\label{fig:texture}
\end{figure*}

\section{Magnon spin polarization}

The spin carried by a magnon follows from the eigenmodes of the full harmonic Hamiltonian. In the basis $\Psi_{\bm q}=(a_{\bm q},b_{\bm q},a_{-\bm q}^{\dagger},b_{-\bm q}^{\dagger})^T$ this Hamiltonian has the form $\mathcal H^{(2)}=\frac12\sum_{\bm q}\Psi_{\bm q}^{\dagger}H(\bm q)\Psi_{\bm q}+\mathrm{const.}$, with
\begin{equation}
H(\bm q)=
\begin{pmatrix}
\varepsilon I_2 & B_{\bm q}\\
B_{\bm q}^{\dagger} & \varepsilon I_2
\end{pmatrix},
\qquad
B_{\bm q}=
\begin{pmatrix}
\Delta & -F_{\bm q}^{*}\\
-F_{-\bm q}^{*} & \Delta
\end{pmatrix}.
\label{eq:Bq}
\end{equation}
Here $I_2$ is the $2\times2$ identity matrix in sublattice space, $\varepsilon=3JS+(K_\theta+K_\varphi)/S$, and $\Delta=(K_\theta-K_\varphi)/S$ measures the difference between the transverse anisotropies. With Fourier phases $e^{i\bm q\cdot\bm r}$, the intersublattice pairing amplitude combines exchange and four-spin contributions,
\begin{equation}
F_{\bm q}
=JS\sum_{\alpha=1}^{3}e^{-i\bm q\cdot\bm\delta_\alpha}
+\frac{\Gamma_{\chi,\bm q}}{S}.
\label{eq:Fq}
\end{equation}
Note that the four-spin term enters only the pair (anomalous) part of the harmonic Hamiltonian, not its number-conserving part. The intersublattice pair term changes the two sublattice occupations equally and therefore conserves the longitudinal spin, the projection of the total spin on $\bm n_0$, at harmonic order. Spin mixing is provided by the same-sublattice pair terms proportional to $\Delta$, which change this spin by $2\hbar$. The magnon spin is then not quantized and must be calculated from the eigenvectors.

The positive-energy modes solve the bosonic eigenvalue problem \cite{Colpa1978}
\begin{equation}
\Sigma_zH(\bm q)u_{n\bm q}
=E_{n\bm q}u_{n\bm q},
\qquad
u_{n\bm q}^{\dagger}\Sigma_z u_{m\bm q}
=\delta_{nm},
\label{eq:dyn}
\end{equation}
where $\Sigma_z=\operatorname{diag}(1,1,-1,-1)$ accounts for the bosonic commutation relations, $\delta_{nm}$ is the Kronecker delta, and we order the bands as $E_{1\bm q}\leq E_{2\bm q}$. We identify the spin $L_{n\bm q}$ of a mode as the change in the total longitudinal angular momentum $J_L=\hbar\sum_{i,\eta}\bm n_0\cdot\bm S_{i\eta}$ on adding one quasiparticle to the Bogoliubov vacuum. Creating an $A$ boson decreases $J_L$ by $\hbar$, creating a $B$ boson increases it by $\hbar$, so that
\begin{equation}
L_{n\bm q}
=\hbar\,u_{n\bm q}^{\dagger}\Lambda_Lu_{n\bm q},
\qquad
\Lambda_L=\operatorname{diag}(-1,1,-1,1),
\label{eq:spin}
\end{equation}
which includes both the particle and the hole components of the eigenvector.

\begin{figure*}[t]
\centering
\includegraphics[width=0.84\textwidth]{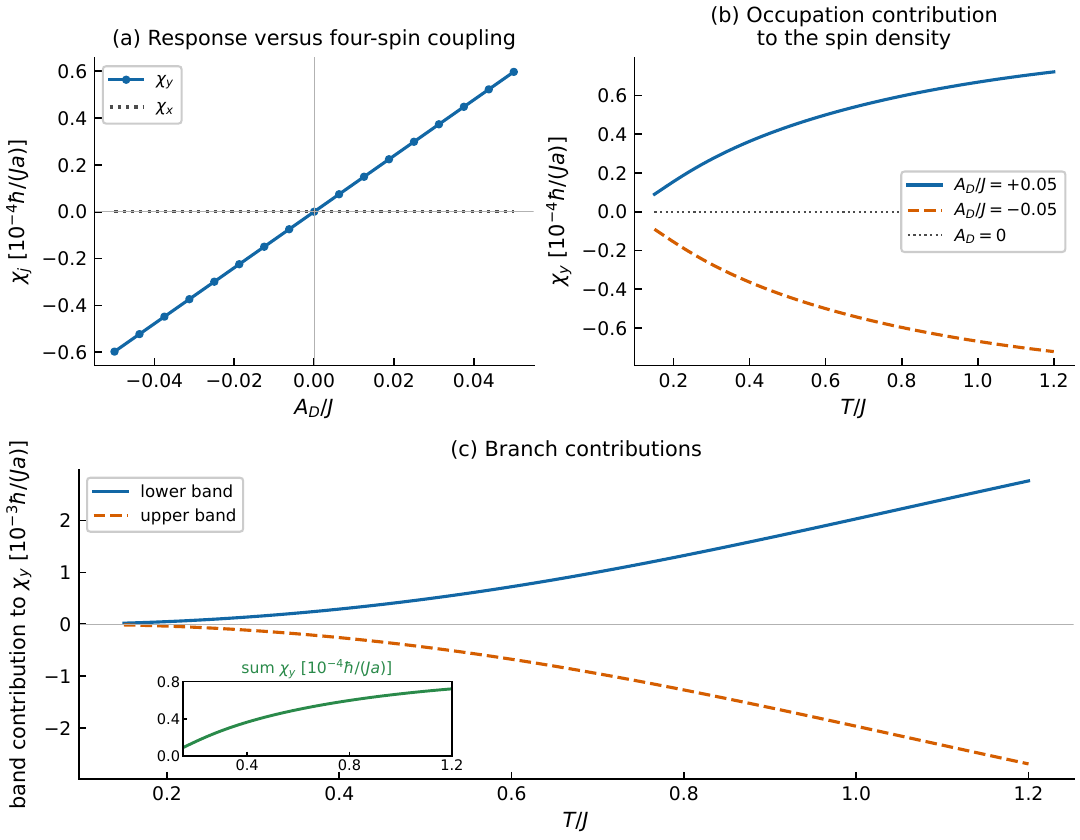}
\caption{
Occupation contribution to the thermal spin density, Eq.~\eqref{eq:response},
for a constant relaxation time $\tau=\hbar/J$; another constant $\tau$
multiplies $\chi_j$ by $\tau J/\hbar$.
(a) Response coefficients versus the four-spin coupling at $T=0.8J$.
The coefficient $\chi_y$ (solid, markers) is linear in $A_D$ over this range and
changes sign with it, while $\chi_x$ (dotted) vanishes by the
symmetry~\eqref{eq:xreflection}.
(b) $\chi_y$ versus temperature for $A_D/J=+0.05$ (solid), $-0.05$ (dashed),
and $0$ (dotted). The response grows as more modes are thermally populated and
is exponentially small below the lowest gap $0.35J$.
(c) Contributions of the lower (solid) and upper (dashed) band to $\chi_y$ at
$A_D/J=+0.05$; the inset shows their sum, the solid curve of panel (b).
Temperature is in energy units ($k_B=1$); the other parameters are those of
Fig.~\ref{fig:texture}.
}
\label{fig:response}
\end{figure*}

The odd parity of the spin follows from a sublattice-exchange relation of the harmonic Hamiltonian. Let $\mathcal P=\operatorname{diag}(\sigma_x,\sigma_x)$, with $\sigma_x$ exchanging the $A$ and $B$ components. The matrices in Eqs.~\eqref{eq:Bq}--\eqref{eq:spin} obey
\begin{equation}
H(-\bm q)=\mathcal P H(\bm q)\mathcal P^\dagger,\;\;
[\mathcal P,\Sigma_z]=0,\;\;
\mathcal P^\dagger\Lambda_L\mathcal P=-\Lambda_L.
\label{eq:symmetry}
\end{equation}
The first two identities map each positive-norm mode at $\bm q$ onto a mode with the same energy at $-\bm q$; for a nondegenerate band $u_{n,-\bm q}=\mathcal P u_{n\bm q}$ up to a phase. The third identity then reverses the spin, and we thus arrive at
\begin{equation}
E_n(-\bm q)=E_n(\bm q),
\qquad
L_n(-\bm q)=-L_n(\bm q).
\label{eq:parities}
\end{equation}
Opposite momenta carry opposite spin within a band whose dispersion remains reciprocal. The transformation $\mathcal P$ relates harmonic matrices; it does not imply inversion symmetry of the microscopic Hamiltonian.

Figure~\ref{fig:texture} shows these relations for the parameters given above. Both branches are gapped and reciprocal [Fig.~\ref{fig:texture}(a)], while the lower-band spin changes sign between opposite momenta [Fig.~\ref{fig:texture}(b)]. The spin vanishes on the line $q_y=0$, where the chiral pairing vanishes, and reaches about $0.12\hbar$ in magnitude elsewhere in the Brillouin zone. Its smooth, nonquantized values reflect an incomplete mixing of opposite-spin modes: the anisotropy mixes them, whereas the chiral pairing splits them, with opposite signs at opposite momenta. At zero four-spin coupling, biaxial anisotropy mixes the two longitudinal-spin sectors equally. The resulting nondegenerate modes have zero longitudinal spin, although their energies are generally different [Fig.~\ref{fig:texture}(c)]. The chiral interaction makes the mixing unequal, with opposite imbalances at opposite momenta. Reversing $A_D$ reverses the spin, and since $H(\bm q;-A_D)=H(-\bm q;A_D)^*$, Eq.~\eqref{eq:parities} implies that the energies do not change. At fixed coupling magnitude, the sign of the four-spin coupling thus controls the spin texture and leaves the energies unchanged; the energies generally depend on the coupling magnitude. (At the zone corners $K$ and $K'$ the two branches are degenerate, and the spin of an individual mode depends on the choice of basis \cite{SupplementalMaterial}.)

In equilibrium the opposite spins at $\bm q$ and $-\bm q$ cancel, because their occupations are equal. A temperature gradient changes the occupations of these counterpropagating modes in opposite directions, so that for equal relaxation times their spin contributions add. For the quadratic Hamiltonian~\eqref{eq:Bq}, the two nondegenerate modes carry opposite longitudinal spins, $L_{2\bm q}=-L_{1\bm q}$ \cite{SupplementalMaterial}. Their thermal contributions do not cancel because their velocities and thermal weights differ.

\section{Thermal spin accumulation}

We calculate the bulk nonequilibrium spin angular momentum per unit layer area induced by a weak temperature gradient through changes in magnon occupations, in the relaxation-time approximation \cite{Li2020,Zhang2020}. The spin density is polarized along the N\'eel axis. Its conversion into a detector signal requires a separate model of spin transfer at the interface. The departure from equilibrium is $\delta n_{n\bm q}=-\tau_{n\bm q}\bm v_{n\bm q}\cdot\nabla T\,\partial_T n_B(E_{n\bm q},T)$, with relaxation time $\tau_{n\bm q}$, group velocity $\bm v_{n\bm q}=\hbar^{-1}\nabla_{\bm q}E_{n\bm q}$, and Bose function $n_B(E,T)=[\exp(E/T)-1]^{-1}$ ($k_B=1$). Summing the spin carried by these occupation changes gives the spin angular momentum per unit area along $\bm n_0$, $\delta s_L=\sum_{j=x,y}\chi_j(-\partial_jT)$, with
\begin{equation}
\chi_j=
\sum_{n=1}^{2}
\int_{\mathrm{BZ}}\frac{\mathrm d^2q}{(2\pi)^2}\,
\tau_{n\bm q}L_{n\bm q}v_{n\bm q,j}
\,\partial_T n_B(E_{n\bm q},T).
\label{eq:response}
\end{equation}
For a constant relaxation time the direction of the response follows from one more relation,
\begin{equation}
H(-q_x,q_y)=H(q_x,q_y)^*.
\label{eq:xreflection}
\end{equation}
Complex conjugation preserves the real matrix $\Lambda_L$, so energy and spin are even under $q_x\to-q_x$, whereas $v_x$ is odd. The integrand of $\chi_x$ therefore cancels and $\chi_x=0$; no such cancellation applies to $\chi_y$.

For a tilt toward the bond $\bm\delta_1$, as considered here, only the $y$ component of the temperature gradient produces a spin accumulation. Rotating the in-plane tilt direction by an angle $\varphi_0$ away from $\bm\delta_1$ leaves the magnitude of the response unchanged and turns it into $(\chi_x,\chi_y)=\chi_0(\sin2\varphi_0,\cos2\varphi_0)$, where $\chi_0$ is the value of $\chi_y$ at $\varphi_0=0$ (we found this numerically for the parameters of Fig.~\ref{fig:response}; the amplitude for an arbitrary tilt direction and the angular check are given in the Supplemental Material \cite{SupplementalMaterial}). An analogous dependence, $\propto\sin^2\theta\cos\theta\,(\sin2\varphi,\cos2\varphi)$ in the polar and azimuthal angles of the magnetization, determines the spiral wave vector induced by quartic asymmetric exchange in ferromagnets \cite{Ado2022}.

For the parameters of Sec.~II at $T=0.8J$ we find $\chi_y=5.97\times10^{-5}\hbar/(Ja)$ for $\tau=\hbar/J$ (convergence checks are in the Supplemental Material \cite{SupplementalMaterial}). Figure~\ref{fig:response}(a) shows that the response is odd in $A_D$, as it should be when the spin reverses at unchanged energies. The response grows with temperature over the range of Fig.~\ref{fig:response}(b), following the thermal weight of the modes in Eq.~\eqref{eq:response}, and is exponentially suppressed below the lowest magnon gap, $0.35J$ at the zone center.

The response also requires the biaxial anisotropy. When $K_\theta=K_\varphi$, the harmonic Hamiltonian conserves $J_L$ and its modes can be labeled by the fixed spin projections $+\hbar$ and $-\hbar$. For a constant relaxation time, the contribution of each spin-labeled branch is an integral of a momentum derivative over the Brillouin zone and vanishes. This spin labeling should be distinguished from ordering the energies as $E_1\leq E_2$: the energy-ordered bands exchange their spin labels at crossings. A momentum-dependent spin removes this cancellation.

Equation~\eqref{eq:response} contains only the contribution from changes in magnon occupations. Interband coherence, which also contributes to the magnon thermal Edelstein effect \cite{Li2020}, is outside this approximation, and a temperature-dependent relaxation time would change the curves in Fig.~\ref{fig:response}(b).

\section{Conclusion}
\label{sec:conclusion}

In summary, chiral four-spin exchange gives the magnons of a compensated honeycomb antiferromagnet a spin that is odd in momentum, provided the N\'eel vector is tilted out of the layer normal and the anisotropy is biaxial. Through intersublattice pairing the interaction acts already on harmonic magnons, complementing the nonlinear magnon processes proposed for a honeycomb ferrimagnet \cite{Murzaliev2025}.

Dzyaloshinskii--Moriya exchange can also produce odd-parity magnon spin and a thermal spin accumulation \cite{Zhang2020}. What distinguishes the four-spin mechanism in our model is its dependence on the orientation of the N\'eel vector: the factor $\cos\theta_0\sin^2\theta_0$ in Eq.~\eqref{eq:gammaexact} and the $(\sin2\varphi_0,\cos2\varphi_0)$ dependence of the response on the in-plane tilt direction. Identifying the coupling in a real material requires comparing this angular dependence with a microscopic model that includes the competing interactions.

Our model is a minimal one. Realizing it in a modified MnPS$_3$ layer requires a structure that permits the four-spin interaction while keeping the two sublattices equivalent, as assumed in Eq.~\eqref{eq:symmetry}. Molecular adsorption has been predicted to alter both the magnetic anisotropy and the Dzyaloshinskii--Moriya interaction of MnPS$_3$ \cite{Wang2022}, which shows how sensitive its magnetic couplings are to structural perturbations. A microscopic calculation for a given structure would have to establish the allowed four-spin term, determine its strength, and assess competing bilinear interactions; the value $A_D=0.05J$ used here specifies the model regime and is not an experimentally established coupling. Our analysis also assumes a locally stable N\'eel state in the harmonic approximation; stability against finite-amplitude, spatially nonuniform configurations requires a separate analysis.

In closing, we comment on the experimental observability. Polarized neutron scattering has resolved opposite magnon polarizations in the ferrimagnet Y$_3$Fe$_5$O$_{12}$ \cite{Nambu2020}, so the magnon spin can be accessed independently of the dispersion. The calculation does not establish the full neutron-scattering intensity. Converting the thermal spin density into an electrical signal requires a model of spin transfer to the detector \cite{Cheng2014}, which remains to be worked out for the present system; electrical measurements of long-distance spin transport \cite{Lebrun2018} and of magnon pseudospin dynamics \cite{Wimmer2020} show that the relevant detection techniques exist for antiferromagnets. At fixed coupling magnitude, reversing $A_D$ leaves the magnon energies unchanged but reverses the mode spin and the occupation-induced thermal spin density. The energies generally depend on the coupling magnitude, so energy measurements alone cannot determine its sign within this model.

\clearpage

\bibliography{references}

\clearpage
\onecolumngrid
\setcounter{section}{0}
\setcounter{equation}{0}
\setcounter{table}{0}
\setcounter{figure}{0}
\renewcommand{\theequation}{S\arabic{equation}}
\renewcommand{\thesection}{S\arabic{section}}
\renewcommand{\thetable}{S\arabic{table}}
\begin{center}
{\large\bfseries Supplemental Material for\\[1mm] ``Odd-parity magnon spin polarization from chiral four-spin exchange in a honeycomb antiferromagnet''}\\[3mm]
B. Murzaliev and M. Titov
\end{center}
\vspace{4mm}

This Supplemental Material contains three derivations for the main text: the harmonic expansion of the four-spin interaction together with a stability bound (Sec.~\ref{sec:harmonic}), the spin carried by a magnon, including its behavior at the zone corners (Sec.~\ref{sec:spin}), and the thermal response with its numerical evaluation (Sec.~\ref{sec:thermal-response}). Throughout we use the compensated N\'eel state and the orientation of the anisotropy axes of the main text.

\section{Harmonic expansion of the spin Hamiltonian}\label{sec:harmonic}

\subsection{Local frames and geometry}

We derive the harmonic Hamiltonian by expanding about the collinear N\'eel state, with spins \(S\mathbf n_0\) on sublattice \(A\) and \(-S\mathbf n_0\) on sublattice \(B\), where \(S\) is the dimensionless spin length. The ordering direction is \(\mathbf n_0=(\sin\theta_0,0,\cos\theta_0)\), with \(\theta_0\) measured from the layer normal. We choose the transverse unit vectors \(\mathbf e_\theta=(\cos\theta_0,0,-\sin\theta_0)\) and \(\mathbf e_\varphi=(0,1,0)\). The local spin axes are then \((\mathbf e_\theta,\mathbf e_\varphi,\mathbf n_0)\) on sublattice \(A\) and \((-\mathbf e_\theta,\mathbf e_\varphi,-\mathbf n_0)\) on sublattice \(B\). Both frames are right-handed, and each local \(z\) axis points along its sublattice's ordered moment. This choice fixes the relative signs in the bosonic expansion below.

We introduce Holstein--Primakoff bosons \(a_i\) and \(b_i\) for the spins on sublattices \(A\) and \(B\), respectively, where \(i\) labels a unit cell. These operators annihilate local spin deviations, and their Hermitian conjugates create them. In the local frames defined above, the leading spin-wave expansion reads
\begin{equation}
\label{eq:hp}
\begin{aligned}
\mathbf S_{iA}
&\simeq
\mathbf n_0(S-a_i^\dagger a_i)
+\sqrt{\frac S2}\left[
\mathbf e_\theta(a_i+a_i^\dagger)
-i\mathbf e_\varphi(a_i-a_i^\dagger)
\right],\\
\mathbf S_{iB}
&\simeq
-\mathbf n_0(S-b_i^\dagger b_i)
+\sqrt{\frac S2}\left[
-\mathbf e_\theta(b_i+b_i^\dagger)
-i\mathbf e_\varphi(b_i-b_i^\dagger)
\right].
\end{aligned}
\end{equation}
The terms linear in bosonic operators describe transverse fluctuations, while the number operators \(a_i^\dagger a_i\) and \(b_i^\dagger b_i\) describe the reduction of the spin component along each local ordering direction. We substitute these expressions into the microscopic Hamiltonian and retain terms up to quadratic order in the bosons.

\subsection{Expansion of the four-spin interaction}

To apply this expansion to the four-spin interaction, we label the six sites of each hexagon by their positions relative to its center \(\mathbf R\). The three \(A\) sites lie at \(\mathbf R+\boldsymbol\delta_\alpha\), and the three \(B\) sites at \(\mathbf R-\boldsymbol\delta_\alpha\), where \(\alpha=1,2,3\). Here \(\boldsymbol\delta_1=a(1,0,0)\) and \(\boldsymbol\delta_{2,3}=a(-1/2,\pm\sqrt3/2,0)\), with \(a\) the nearest-neighbor distance. The spin projections entering the interaction are
\begin{equation}
\label{eq:projections}
p_{A\alpha}(\mathbf R)
=\frac{\boldsymbol\delta_\alpha\cdot\mathbf S_A(\mathbf R+\boldsymbol\delta_\alpha)}{S},
\qquad
p_{B\alpha}(\mathbf R)
=\frac{\boldsymbol\delta_\alpha\cdot\mathbf S_B(\mathbf R-\boldsymbol\delta_\alpha)}{S}.
\end{equation}
Here \(\mathbf S_\eta(\mathbf r)\) denotes the spin operator \(\mathbf S_{i\eta}\) at the position \(\mathbf r\) of sublattice \(\eta\). The same vector \(+\boldsymbol\delta_\alpha\) defines both projections, including the \(B\) site at \(\mathbf R-\boldsymbol\delta_\alpha\). This convention fixes the sign of the four-spin interaction, which we write explicitly as
\begin{equation}
\label{eq:microscopic-chiral}
\mathcal H_\chi=\frac{A_D}{a^4}\sum_{\mathbf R}\Biggl[
\Bigl(\prod_{\alpha=1}^{3}p_{A\alpha}(\mathbf R)\Bigr)\Bigl(\sum_{\beta=1}^{3}p_{B\beta}(\mathbf R)\Bigr)
-\Bigl(\prod_{\alpha=1}^{3}p_{B\alpha}(\mathbf R)\Bigr)\Bigl(\sum_{\beta=1}^{3}p_{A\beta}(\mathbf R)\Bigr)\Biggr].
\end{equation}
Equation~\eqref{eq:microscopic-chiral} is Eq.~(2) of the main text. Here the sum over \(\mathbf R\) includes each hexagon once, and \(A_D\) is the four-spin coupling constant, with units of energy. Each projection has units of length, so the factor \(a^{-4}\) makes the product of four projections dimensionless.

The expansion of these projections can be written compactly using the geometric coefficients \(x_\alpha=\boldsymbol\delta_\alpha\cdot\mathbf n_0\) and \(z_\alpha=\boldsymbol\delta_\alpha\cdot(\mathbf e_\theta-i\mathbf e_\varphi)\). The real coefficient \(x_\alpha\) describes the projection along the ordering direction, while the complex coefficient \(z_\alpha\) combines the two transverse projections. For a fixed hexagon, we write \(a_\alpha\) and \(b_\alpha\) for the bosonic operators at \(\mathbf R+\boldsymbol\delta_\alpha\) and \(\mathbf R-\boldsymbol\delta_\alpha\), respectively, and suppress \(\mathbf R\). Substituting the spin-wave expansion gives
\begin{equation}
\label{eq:projection-expansion}
p_{A\alpha}
\simeq x_\alpha
+\frac{z_\alpha a_\alpha+z_\alpha^{*}a_\alpha^\dagger}{\sqrt{2S}}
-\frac{x_\alpha}{S}a_\alpha^\dagger a_\alpha,
\qquad
p_{B\alpha}
\simeq -x_\alpha
-\frac{z_\alpha^{*}b_\alpha+z_\alpha b_\alpha^\dagger}{\sqrt{2S}}
+\frac{x_\alpha}{S}b_\alpha^\dagger b_\alpha,
\end{equation}
where the asterisk denotes complex conjugation. The opposite constant terms reflect the antiparallel ordered moments. Since \(\sum_{\alpha=1}^{3}\boldsymbol\delta_\alpha=0\), the geometric coefficients obey \(\sum_\alpha x_\alpha=\sum_\alpha z_\alpha=0\); these identities determine which terms survive in the expansion of the four-spin interaction.

The classical four-spin energy vanishes because the sum of the three projections on either sublattice is zero in the N\'eel state. To establish stationarity, we also evaluate the terms linear in the bosons. Let \(p_{\eta\alpha}^{(1)}(\mathbf R)\) denote the linear part of the projection on sublattice \(\eta=A,B\), and define \(X=x_1x_2x_3\). The classical products of the three projections are \(X\) on sublattice \(A\) and \(-X\) on sublattice \(B\), giving
\begin{equation}
\label{eq:linear-chiral}
\mathcal H_\chi^{(1)}
=\frac{A_D X}{a^4}
\sum_{\mathbf R}\sum_{\alpha=1}^{3}
\left[
p_{A\alpha}^{(1)}(\mathbf R)
+p_{B\alpha}^{(1)}(\mathbf R)
\right].
\end{equation}
In a bulk system with periodic boundaries, each physical site belongs to three hexagons and appears once with each vertex label \(\alpha\). Its linear contributions therefore sum to a coefficient proportional to \(\sum_\alpha z_\alpha\) or its complex conjugate, both of which vanish. The cancellation occurs after summing over hexagons and establishes that the four-spin interaction produces no bulk torque on the uniform N\'eel state.

At quadratic order, the four-spin interaction contains contributions from the boson number terms in the projections and from products of two transverse fluctuations. The number terms cancel after summing over hexagons: at each physical site, their coefficient is proportional to \(\sum_\alpha x_\alpha=0\). To express the remaining contribution, we define \(f_\alpha=\prod_{\gamma\ne\alpha}x_\gamma\), where \(\gamma=1,2,3\) labels the vertices on one sublattice. Thus, \(f_\alpha\) is the product of the two classical projections left when the projection at vertex \(\alpha\) is replaced by its linear fluctuation. The quadratic interaction becomes
\begin{equation}
\label{eq:quadratic-chiral-real}
\mathcal H_\chi^{(2)}
=\frac{A_D}{a^4}
\sum_{\mathbf R}\sum_{\alpha,\beta=1}^{3}
(f_\alpha-f_\beta)\,
p_{A\alpha}^{(1)}(\mathbf R)\,
p_{B\beta}^{(1)}(\mathbf R).
\end{equation}
The two coefficients arise from expanding the three-spin product on sublattice \(A\) or \(B\), respectively; their relative minus sign follows from the microscopic interaction. The surviving terms therefore couple transverse fluctuations on opposite sublattices, with a geometric coefficient that is antisymmetric under \(\alpha\leftrightarrow\beta\). This antisymmetry determines which bosonic terms survive in momentum space.

To see which terms survive, we Fourier transform the bosonic operators using the physical positions of the spins,
\begin{equation}
\label{eq:fourier}
a_{\mathbf r}=\frac{1}{\sqrt{N_c}}\sum_{\mathbf q}
e^{i\mathbf q\cdot\mathbf r}a_{\mathbf q},
\qquad
b_{\mathbf r}=\frac{1}{\sqrt{N_c}}\sum_{\mathbf q}
e^{i\mathbf q\cdot\mathbf r}b_{\mathbf q},
\end{equation}
where \(N_c\) is the number of unit cells, \(\mathbf q\) runs over the first Brillouin zone, and \(\mathbf r\) belongs to the corresponding sublattice; \(a_{\mathbf r}\) and \(b_{\mathbf r}\) are the bosons \(a_i\) and \(b_i\) (or \(a_\alpha\) and \(b_\alpha\)) at that site. Substituting the linear projections into \(\mathcal H_\chi^{(2)}\) produces both number-conserving terms, such as \(a_{\mathbf q}b_{\mathbf q}^{\dagger}\), and pair terms, such as \(a_{\mathbf q}b_{-\mathbf q}\). The coefficient of the former is proportional to
\begin{equation}
\label{eq:normal-cancellation}
\sum_{\alpha,\beta=1}^{3}
(f_\alpha-f_\beta)\,z_\alpha z_\beta\,
e^{i\mathbf q\cdot(\boldsymbol\delta_\alpha+\boldsymbol\delta_\beta)}
=0.
\end{equation}
This cancellation follows because \(f_\alpha-f_\beta\) is antisymmetric under exchange of the vertex indices, whereas the remaining factors are symmetric. The Hermitian-conjugate coefficient vanishes for the same reason. Thus, at harmonic order, the four-spin interaction contributes only terms that create or annihilate a pair of magnons on opposite sublattices.

The surviving pair terms define the chiral contribution to the harmonic Hamiltonian,
\begin{equation}
\label{eq:chiral-pairing}
\mathcal H_\chi^{(2)}
=-\frac{1}{S}\sum_{\mathbf q}
\left[
\Gamma_{\chi,\mathbf q}\,a_{\mathbf q}b_{-\mathbf q}
+\Gamma_{\chi,\mathbf q}^{*}\,
a_{\mathbf q}^{\dagger}b_{-\mathbf q}^{\dagger}
\right],
\end{equation}
where the pairing amplitude
\begin{equation}
\label{eq:gamma-definition}
\Gamma_{\chi,\mathbf q}
=\frac{A_D}{2a^4}
\sum_{\alpha,\beta=1}^{3}
(f_\alpha-f_\beta)\,
z_\alpha z_\beta^{*}\,
e^{i\mathbf q\cdot(\boldsymbol\delta_\alpha+\boldsymbol\delta_\beta)}
\end{equation}
has units of energy. The factor \(1/(2S)\) comes from the two linear spin projections, and the overall minus sign follows from the transverse expansion on sublattice \(B\). The product \(z_\alpha z_\beta^{*}\) has an antisymmetric part that survives the vertex sum and determines the pairing amplitude. Exchanging \(\alpha\) and \(\beta\) after complex conjugation gives \(\Gamma_{\chi,-\mathbf q}=-\Gamma_{\chi,\mathbf q}^{*}\). Consequently, its real part is odd in momentum, while its imaginary part is even.

For the honeycomb geometry and local axes specified above, the vertex sum evaluates to
\begin{equation}
\label{eq:gamma-explicit}
\Gamma_{\chi,\mathbf q}
=-\frac{3\sqrt3}{4}\,
A_D\cos\theta_0\sin^2\theta_0\,
e^{iaq_x/2}
\sin\!\left(\frac{\sqrt3\,a q_y}{2}\right),
\end{equation}
where \(q_x\) and \(q_y\) are the Cartesian components of the in-plane wave vector; this is Eq.~(4) of the main text. Near the Brillouin-zone center, this expression becomes
\begin{equation}
\label{eq:gamma-small-q}
\Gamma_{\chi,\mathbf q}
=-\frac{9}{8}\,
A_D a\cos\theta_0\sin^2\theta_0\,q_y
+O\!\left(A_Da^2|\mathbf q|^2\right).
\end{equation}
The leading contribution is therefore real and linear in \(q_y\). The angular factor shows that the chiral contribution to harmonic pairing requires a tilted ordering direction: it vanishes for moments aligned with the layer normal (\(\theta_0=0\)) or with the in-plane \(x\) axis (\(\theta_0=\pi/2\)). These zeros concern the quadratic expansion about the chosen N\'eel state; the microscopic four-spin interaction can still contribute at higher orders in the fluctuations.

\subsection{Quadratic Hamiltonian}

It remains to combine this contribution with the nearest-neighbor exchange and the transverse single-ion anisotropy,
\begin{equation}
\label{eq:exchange-anisotropy}
\mathcal H_0
=J\sum_{\langle ij\rangle}\mathbf S_{iA}\cdot\mathbf S_{jB}
+\sum_{i,\eta=A,B}\left[
K_\theta\left(\frac{\mathbf S_{i\eta}\cdot\mathbf e_\theta}{S}\right)^2
+K_\varphi\left(\frac{\mathbf S_{i\eta}\cdot\mathbf e_\varphi}{S}\right)^2
\right].
\end{equation}
Here \(\langle ij\rangle\) denotes nearest-neighbor pairs, with each bond counted once; \(J>0\) is the antiferromagnetic exchange constant, and \(K_\theta,K_\varphi>0\) are anisotropy energies that penalize spin components along the two transverse axes. The factors of \(S\) normalize each spin component before squaring. Up to an additive constant, the complete quadratic Hamiltonian is
\begin{equation}
\label{eq:quadratic-total}
\mathcal H^{(2)}
=\sum_{\mathbf q}\Bigl\{
\varepsilon\left(a_{\mathbf q}^{\dagger}a_{\mathbf q}
+b_{\mathbf q}^{\dagger}b_{\mathbf q}\right)
-\left[F_{\mathbf q}a_{\mathbf q}b_{-\mathbf q}
+\mathrm{H.c.}\right]
+\frac{\Delta}{2}
\left[a_{\mathbf q}a_{-\mathbf q}
+b_{\mathbf q}b_{-\mathbf q}
+\mathrm{H.c.}\right]\Bigr\},
\end{equation}
where \(\mathrm{H.c.}\) denotes the Hermitian conjugate of the preceding terms. The diagonal energy \(\varepsilon\), the same-sublattice pairing amplitude \(\Delta\), and the intersublattice pairing amplitude \(F_{\mathbf q}\) are
\begin{equation}
\label{eq:harmonic-coefficients}
\varepsilon=3JS+\frac{K_\theta+K_\varphi}{S},
\qquad
\Delta=\frac{K_\theta-K_\varphi}{S},
\qquad
F_{\mathbf q}
=JS\sum_{\alpha=1}^{3}e^{-i\mathbf q\cdot\boldsymbol\delta_\alpha}
+\frac{\Gamma_{\chi,\mathbf q}}{S}.
\end{equation}
The three nearest neighbors give the exchange contribution \(3JS\) to \(\varepsilon\). The difference between the transverse anisotropies produces same-sublattice pairing, which vanishes when \(K_\theta=K_\varphi\).

To write the quadratic Hamiltonian in matrix form, we collect the annihilation and creation operators into the Nambu vector \(\Psi_{\mathbf q}=(a_{\mathbf q},b_{\mathbf q},a_{-\mathbf q}^{\dagger},b_{-\mathbf q}^{\dagger})^{T}\), where \(T\) denotes transpose. The Hamiltonian then takes the bosonic Bogoliubov--de Gennes form
\begin{equation}
\label{eq:bdg}
\mathcal H^{(2)}
=\frac12\sum_{\mathbf q}
\Psi_{\mathbf q}^{\dagger}H(\mathbf q)\Psi_{\mathbf q}
+\mathrm{const.},
\qquad
H(\mathbf q)
=
\begin{pmatrix}
\varepsilon I_2 & B_{\mathbf q}\\
B_{\mathbf q}^{\dagger} & \varepsilon I_2
\end{pmatrix},
\qquad
B_{\mathbf q}
=
\begin{pmatrix}
\Delta & -F_{\mathbf q}^{*}\\
-F_{-\mathbf q}^{*} & \Delta
\end{pmatrix}.
\end{equation}
Equation~\eqref{eq:bdg} is Eq.~(5) of the main text. Here \(I_2\) is the two-dimensional identity matrix acting on the sublattice indices, and \(B_{\mathbf q}\) contains the pair-creation amplitudes. Its diagonal entries describe pairing within each sublattice, while its off-diagonal entries describe pairing between sublattices. The relation \(B_{-\mathbf q}^{T}=B_{\mathbf q}\) expresses the symmetry under exchange of the two bosons in a pair. The factor \(1/2\) compensates for the duplication introduced by including both annihilation and creation operators in \(\Psi_{\mathbf q}\).

The magnon energies follow from the equations of motion, which must retain the bosonic commutation relations. In the chosen Nambu basis, these relations are encoded by \(\Sigma_z=\operatorname{diag}(1,1,-1,-1)\): the opposite signs distinguish annihilation from creation operators. For a locally stable state, the two physical magnon branches are obtained from
\begin{equation}
\label{eq:magnon-modes}
\Sigma_z H(\mathbf q)u_{n\mathbf q}
=E_{n\mathbf q}u_{n\mathbf q},
\qquad
u_{n\mathbf q}^{\dagger}\Sigma_z u_{m\mathbf q}
=\delta_{nm},
\end{equation}
where \(n,m=1,2\) label the positive-energy modes (we also write \(E_n(\mathbf q)\equiv E_{n\mathbf q}\) and \(L_n(\mathbf q)\equiv L_{n\mathbf q}\)), \(E_{n\mathbf q}>0\) is the magnon energy, and \(u_{n\mathbf q}\) is the four-component vector of mode amplitudes in the Nambu basis. Here \(\delta_{nm}\) is the Kronecker delta. The normalization with \(\Sigma_z\) ensures that the corresponding quasiparticle operators obey canonical bosonic commutation relations. The excitation energies are therefore eigenvalues of \(\Sigma_z H(\mathbf q)\); the ordinary eigenvalues of the Hermitian matrix \(H(\mathbf q)\) instead determine whether fluctuations increase the energy of the reference N\'eel state. We examine this stability condition below.

\subsection{Harmonic stability}

Local harmonic stability requires \(H(\mathbf q)\) to be positive definite at every momentum, so that any nonzero fluctuation raises the energy of the reference state. We denote its smallest ordinary eigenvalue by \(\lambda_{\min}(H(\mathbf q))\). For the block structure derived above, this eigenvalue equals \(\varepsilon\) minus the largest singular value of \(B_{\mathbf q}\), namely the square root of the largest eigenvalue of \(B_{\mathbf q}^{\dagger}B_{\mathbf q}\). The latter singular value is bounded by \(|\Delta|+\max(|F_{\mathbf q}|,|F_{-\mathbf q}|)\). Using \(\bigl|\sum_{\alpha}e^{-i\mathbf q\cdot\boldsymbol\delta_\alpha}\bigr|\leq3\) and the explicit expression for \(\Gamma_{\chi,\mathbf q}\), we therefore obtain
\begin{equation}
\label{eq:stability}
\lambda_{\min}(H(\mathbf q))
\geq \varepsilon-|\Delta|
-\max\!\left(|F_{\mathbf q}|,|F_{-\mathbf q}|\right)
\geq
\frac{2\min(K_\theta,K_\varphi)}{S}
-\frac{3\sqrt3\,|A_D\cos\theta_0|\sin^2\theta_0}{4S}.
\end{equation}
A strictly positive right-hand side of Eq.~\eqref{eq:stability} establishes stability throughout the Brillouin zone without requiring a momentum scan. This is a sufficient condition: if the bound becomes zero or negative, the eigenvalues of \(H(\mathbf q)\) must be examined directly before concluding that the state is unstable.

For the illustrative parameters used in the main text, \(S=5/2\), \(\theta_0=8^\circ\), \(K_\theta=0.02J\), \(K_\varphi=0.01J\), and \(A_D=0.05J\), the bound gives
\[
\lambda_{\min}(H(\mathbf q))\geq 7.50\times10^{-3}J>0
\]
throughout the Brillouin zone. The compensated Néel state is therefore locally stable within the harmonic approximation, and the corresponding magnon energies are real and positive. This result justifies using the quadratic Hamiltonian to calculate the mode spin and thermal response for these parameters. It does not establish that the Néel state is the global ground state of the full spin Hamiltonian, since competing configurations at finite deviations from this state are not tested by the harmonic expansion.

\section{Spin carried by a magnon}
\label{sec:spin}

We define the longitudinal spin angular momentum as the projection of the total spin along the Néel direction \(\mathbf n_0\),
\[
J_L=\hbar\sum_{i,\eta=A,B}\mathbf n_0\cdot\mathbf S_{i\eta},
\]
where \(\hbar\) restores the units of angular momentum for the dimensionless spin operators. The classical contributions from the two sublattices cancel because they contain equal numbers of spins with opposite ordered moments. Substituting the Holstein–Primakoff representation gives
\[
J_L
=\hbar\sum_i\left(-a_i^\dagger a_i+b_i^\dagger b_i\right)
=\hbar\sum_{\mathbf q}
\left(-a_{\mathbf q}^\dagger a_{\mathbf q}
+b_{\mathbf q}^\dagger b_{\mathbf q}\right).
\]
An \(A\)-sublattice boson therefore decreases \(J_L\) by \(\hbar\), whereas a \(B\)-sublattice boson increases it by \(\hbar\). Both signs refer to the same global direction \(\mathbf n_0\). To obtain the spin carried by a magnon eigenmode, we evaluate the change in this operator’s expectation value when one Bogoliubov quasiparticle is added to the spin-wave vacuum.

Let \(\beta_{n\mathbf q}^{\dagger}\) create a quasiparticle in the normalized mode \(u_{n\mathbf q}\), and let \(|0\rangle\) denote the Bogoliubov vacuum, which is annihilated by every \(\beta_{n\mathbf q}\). The spin carried by one magnon is
\[
L_n(\mathbf q)
=\langle1_{n\mathbf q}|J_L|1_{n\mathbf q}\rangle
-\langle0|J_L|0\rangle,
\qquad
|1_{n\mathbf q}\rangle=\beta_{n\mathbf q}^{\dagger}|0\rangle.
\]
To evaluate this difference, we write the mode vector as \(u_{n\mathbf q}=(u_A,u_B,v_A,v_B)^T\), suppressing the mode and momentum indices on its components. The amplitudes \(u_\eta\) and \(v_\eta\) multiply annihilation and creation operators, respectively, in the Nambu basis. Adding one quasiparticle changes the total boson occupation on sublattice \(\eta\) by \(|u_\eta|^2+|v_\eta|^2\). Hence
\[
L_n(\mathbf q)
=\hbar\left(-|u_A|^2+|u_B|^2-|v_A|^2+|v_B|^2\right)
=\hbar\,u_{n\mathbf q}^{\dagger}\Lambda_Lu_{n\mathbf q},
\qquad
\Lambda_L=\operatorname{diag}(-1,1,-1,1).
\]
The matrix \(\Lambda_L\) assigns the longitudinal spin sign to each sublattice in both Nambu sectors. The creation-sector weights enter with the same sublattice signs as the annihilation-sector weights because both contribute to the change in occupation. There is no additional factor \(1/2\): the expression already counts the full occupation change caused by one normalized quasiparticle.

At the zone corners \(K\) and \(K'=-K\), the two positive-energy modes are degenerate, and the spin expectation of an arbitrarily chosen eigenvector is basis dependent. The trace and the eigenvalues of the spin operator projected onto the degenerate subspace are basis independent: the projected spin matrix is \(\hbar U^\dagger\Lambda_LU\), where the columns of the \(4\times2\) matrix \(U\) span the degenerate positive-energy subspace and \(U^\dagger\Sigma_zU=I_2\), and both its eigenvalues and its trace are invariant under unitary changes of basis within the degenerate subspace. In the nondegenerate neighborhood the mode spin is well defined, but its limit at a zone corner depends on the direction of approach. For example, approaching \(K=(0,4\pi/3\sqrt3a)\) along \(q_x=0\) from inside the first Brillouin zone gives \(L_1\simeq0.107\hbar\) and \(L_2\simeq-0.107\hbar\) for the parameters of the main text. The isolated degeneracy points have zero measure in the thermal-response integral of Sec.~\ref{sec:thermal-response}.

The harmonic matrix~\eqref{eq:bdg} can be diagonalized in closed form. We define
\[
t_{\mathbf q}=JS\sum_\alpha e^{-i\mathbf q\cdot\boldsymbol\delta_\alpha},\qquad
g_{\mathbf q}=\Gamma_{\chi,\mathbf q}/S,\qquad
d_{\mathbf q}=\operatorname{Re}\bigl(t_{\mathbf q}^{*}g_{\mathbf q}\bigr),\qquad
R_{\mathbf q}=\sqrt{d_{\mathbf q}^{2}+\Delta^{2}|t_{\mathbf q}|^{2}},
\]
so that \(F_{\mathbf q}=t_{\mathbf q}+g_{\mathbf q}\) and \(F_{-\mathbf q}=t_{\mathbf q}^{*}-g_{\mathbf q}^{*}\). For a stable matrix, \(R_{\mathbf q}>0\), and the energy ordering \(E_1\leq E_2\), one obtains
\begin{equation}
\label{eq:closed-energy}
E_{1,2}^{2}(\mathbf q)=\varepsilon^{2}-\Delta^{2}-|t_{\mathbf q}|^{2}-|g_{\mathbf q}|^{2}\mp2R_{\mathbf q},
\qquad
L_1(\mathbf q)=-\hbar\frac{d_{\mathbf q}}{R_{\mathbf q}},\qquad L_2(\mathbf q)=+\hbar\frac{d_{\mathbf q}}{R_{\mathbf q}}.
\end{equation}
The squared energies are the eigenvalues of \(\varepsilon^2I_2-B_{\mathbf q}B_{\mathbf q}^{\dagger}\). The matrices \(B_{\mathbf q}B_{\mathbf q}^{\dagger}\) and \(B_{\mathbf q}^{\dagger}B_{\mathbf q}\) have opposite diagonal differences and equal off-diagonal magnitudes. Their nondegenerate singular modes therefore have opposite sublattice imbalances in the particle (annihilation) and hole (creation) sectors of the Nambu vector. Bosonic normalization fixes the difference of the particle and hole weights to unity, giving the spin formula above.

These expressions show explicitly that \(L_2=-L_1\), that the spin is odd in \(A_D\), and that the energies are even in \(A_D\) but depend on its magnitude. They also explain why the ratio \(d_{\mathbf q}/R_{\mathbf q}\) has a direction-dependent limit as \(t_{\mathbf q}\to0\) at a zone corner. Direct diagonalization at 100 random momenta agrees with these expressions within approximately \(10^{-14}J\) for energies and \(10^{-11}\hbar\) for spins.

\section{Thermal spin accumulation}
\label{sec:thermal-response}

We evaluate the contribution to the longitudinal spin density caused by a redistribution of magnon occupations under a temperature gradient. The spectrum and mode spin are taken from the equilibrium harmonic Hamiltonian. We assume well-defined magnon quasiparticles and a temperature profile that varies slowly on the scale of their mean free path. In the relaxation-time approximation, balancing the drift of the local Bose distribution against relaxation gives
\begin{equation}
\label{eq:thermal-occupation}
\delta n_{n\mathbf q}
=-\tau\,\mathbf v_{n\mathbf q}\cdot\nabla T\,
\frac{\partial n_B(E_{n\mathbf q},T)}{\partial T},
\qquad
n_B(E,T)=\frac{1}{e^{E/T}-1}.
\end{equation}
Here \(\tau\) is a common relaxation time for the two modes, and \(\mathbf v_{n\mathbf q}=\hbar^{-1}\nabla_{\mathbf q}E_{n\mathbf q}\) is their group velocity. We set \(k_B=1\) and take the temperature derivative at fixed energy. This treatment describes the occupation contribution to the response; it does not include coherences induced between different magnon modes.

The resulting change in longitudinal spin angular momentum per unit area is
\begin{equation}
\label{eq:thermal-spin-density}
\delta s_L
=\sum_{n=1}^{2}\int_{\mathrm{BZ}}\frac{\mathrm d^2q}{(2\pi)^2}
L_n(\mathbf q)\,\delta n_{n\mathbf q}
=\sum_{j=x,y}\chi_j(-\partial_jT),
\end{equation}
where the integral covers the first Brillouin zone and \(\chi_j\) is the response to a temperature gradient along direction \(j\). Substitution of Eq.~\eqref{eq:thermal-occupation} yields
\begin{equation}
\label{eq:thermal-response}
\chi_j
=\tau\sum_{n=1}^{2}\int_{\mathrm{BZ}}\frac{\mathrm d^2q}{(2\pi)^2}
L_n(\mathbf q)\,v_{n\mathbf q,j}
\frac{E_{n\mathbf q}}{T^2}
n_B(E_{n\mathbf q},T)
\bigl[1+n_B(E_{n\mathbf q},T)\bigr],
\end{equation}
with \(v_{n\mathbf q,j}\) the corresponding velocity component. For a common constant relaxation time this is Eq.~(11) of the main text. Reciprocal energies imply opposite velocities at opposite momenta. Together with the odd mode spin, this makes the product \(L_n v_{n\mathbf q,j}\) even in momentum, so opposite momenta can contribute with the same sign. The value of the response nevertheless depends on the weighted sum over both branches and the full Brillouin zone.

The integration measure defines a response per physical layer area. We express energies and temperatures in units of \(J\), lengths in units of \(a\), and response coefficients in units of \(\hbar/(Ja)\). Since Eq.~\eqref{eq:thermal-response} is proportional to \(\tau\), the reported amplitudes use \(\tau J/\hbar=1\) as a normalization. This choice does not specify a microscopic scattering time: results for a chosen constant \(\tau\) are obtained by multiplying the reported values by \(\tau J/\hbar\).

For the numerical integration, we sample the midpoints of an \(N\times N\) subdivision of the reciprocal cell spanned by \(\mathbf G_{1,2}=(2\pi/3a,\mp2\pi/\sqrt3a)\). Each point has weight \(1/(A_cN^2)\), where \(A_c=3\sqrt3a^2/2\) is the real-space unit-cell area. This weight preserves the continuum normalization in Eq.~\eqref{eq:thermal-response}. With Fourier phases defined by the physical positions, the matrix \(H(\mathbf q)\) is gauge covariant under reciprocal-lattice translations, whereas the energies, the mode spin, and the integration kernel are periodic, so the reciprocal cell may replace the first Brillouin zone. The velocities are obtained from centered differences of the mode energies with momentum step \(2\times10^{-5}/a\), evaluated independently of the integration mesh. Table~\ref{tab:convergence} shows the convergence of \(\chi_y\) with the mesh for the parameters of the main text at \(T=0.8J\). The converged value is \(\chi_y\simeq5.97\times10^{-5}\hbar/(Ja)\) for \(\tau=\hbar/J\).

\begin{table}[h]
\caption{Convergence of \(\chi_y\) with the mesh, for the parameters of the main text at \(T=0.8J\) and \(\tau=\hbar/J\).}
\label{tab:convergence}
\begin{ruledtabular}
\begin{tabular}{cc}
Integration grid & \(\chi_y\) [\(10^{-5}\hbar/(Ja)\)]\\
\hline
\(84\times84\) & 5.96924\\
\(168\times168\) & 5.96955\\
\(336\times336\) & 5.96955\\
\end{tabular}
\end{ruledtabular}
\end{table}

\subsection{Dependence on the in-plane tilt direction}
\label{sec:angular}

For a tilt rotated by an azimuthal angle \(\varphi_0\) about the layer normal, the N\'eel axis and the rotated local axes are
\[
\mathbf n_0=(\sin\theta_0\cos\varphi_0,\sin\theta_0\sin\varphi_0,\cos\theta_0),\quad
\mathbf e_\theta=(\cos\theta_0\cos\varphi_0,\cos\theta_0\sin\varphi_0,-\sin\theta_0),\quad
\mathbf e_\varphi=(-\sin\varphi_0,\cos\varphi_0,0),
\]
with the anisotropy axes rotated together with the N\'eel axis. Only the geometric coefficients \(x_\alpha\) and \(z_\alpha\) change, and the vertex sum~\eqref{eq:gamma-definition} gives
\begin{equation}
\label{eq:gamma-phi}
\Gamma_{\chi,\mathbf q}
=-\frac{3i}{4}A_D\cos\theta_0\sin^2\theta_0
\sum_{\alpha=1}^{3}\sin[2(\varphi_0-\varphi_\alpha)]\,e^{-i\mathbf q\cdot\boldsymbol\delta_\alpha},
\end{equation}
with \(\varphi_\alpha=0,\,2\pi/3,\,-2\pi/3\) the polar angles of \(\boldsymbol\delta_\alpha\). For \(\varphi_0=0\) this reduces to Eq.~\eqref{eq:gamma-explicit}. As an angular check, we computed the response with this amplitude for \(0\le\varphi_0\le90^\circ\) and the other parameters of the main text: it follows \((\chi_x,\chi_y)=\chi_0(\sin2\varphi_0,\cos2\varphi_0)\), with \(\chi_0\) the value of \(\chi_y\) at \(\varphi_0=0\), to a relative deviation below \(2\times10^{-7}\).

\end{document}